\documentclass[usenatbib,a4paper,times]{mnras}

\usepackage{aas_macros}
\usepackage{amsmath}
\usepackage{amssymb}
\usepackage{bm}
\usepackage{breakurl}
\usepackage{graphicx}
\usepackage{natbib}
\usepackage{times}
\usepackage{textcomp}
\usepackage{url}

\newcommand{\st}{\mathrm{St}}
\renewcommand{\sun}{\mathrm{M}_{\odot}}
\renewcommand{\earth}{\mathrm{M}_{\oplus}}
\newcommand{\mcfost}{\textsc{mcfost}}
\renewcommand{\phantom}{\textsc{phantom}}

\title[Super-Earths in the TW~Hya disc]{Super-Earths in the TW~Hya disc}

\author[Mentiplay, Price \& Pinte]{\parbox{\textwidth}{Daniel
   Mentiplay$^{1}$\thanks{daniel.mentiplay@monash.edu}, Daniel J. Price$^{1}$
   and Christophe Pinte$^{1,2}$}\\
   $^{1}$Monash Centre for Astrophysics (MoCA) and School of Physics and
   Astronomy, Monash University, Clayton Vic 3800, Australia \\
   $^{2}$Univ. Grenoble Alpes, CNRS, IPAG, F-38000 Grenoble, France}

\pagerange{\pageref{firstpage}--\pageref{lastpage}} \pubyear{2018}

\date{}

\begin{document}
\label{firstpage}
\bibliographystyle{mnras}
\maketitle

\begin{abstract}

We test the hypothesis that the sub-millimetre thermal emission and scattered
light gaps seen in recent observations of TW~Hya are caused by planet-disc
interactions. We perform global three-dimensional dusty smoothed particle
hydrodynamics simulations, comparing synthetic observations of our models with
dust thermal emission, CO emission and scattered light observations.  We find
that the dust gaps observed at 24~au and 41~au can be explained by two
super-Earths ($\sim$~4~$\earth{}$). A planet of approximately Saturn-mass can
explain the CO emission and the depth and width of the gap seen in scattered
light at 94~au. Our model produces a prominent spiral arm while there are only
hints of this in the data. To avoid runaway growth and migration of the planets
we require a disc mass of $\lesssim 10^{-2}\,\sun{}$ in agreement with CO
observations but 10--100 times lower than the estimate from HD line emission.

\end{abstract}


\section{Introduction}

TW~Hya, the nearest gas-rich protoplanetary disc, was recently imaged by ALMA at
$870\,\mu$m \citep{andrews:2016}. These observations of thermal emission from
$\sim$100$\mu$m dust in the midplane show a series of stunning axisymmetric
gaps. At just $60$~pc \citep{gaia-collaboration:2018} TW~Hya presents a unique
opportunity to observe planet formation on our doorstep. Being a member of the
3--20~Myr old \citep{barrado-y-navascues:2006} TW~Hya association means TW~Hya
is older than the typical disc lifetime of $\sim$~3~Myr \citep{haisch:2001}
implying that planet formation should almost be complete.

\citet{van-boekel:2017} observed TW~Hya in polarized scattered light using the
Spectro-Polarimetric High-contrast Exoplanet REsearch (SPHERE) instrument on the
Very Large Telescope. Scattered light observations trace the small grains in the
upper layers of the gas disc. These grains are tightly coupled to the gas via
drag. Of the two main gaps in the sub-mm dust emission (at 24~au and 41~au) only
the inner gap is observed in the scattered light image.

Estimates of the gas mass in TW~Hya vary over several orders of magnitude.
\citet{thi:2010} use radiative transfer modelling of CO emission to infer a gas
mass $(0.5\textendash 5)\times10^{-3}~\sun{}$. Whereas \citet{bergin:2013} use
hydrogen deuteride (HD) observations to infer a disc mass $>0.05\,\sun{}$. At
this mass the self-gravity of the disc is significant and gravitational
instability may lead to disc fragmentation \citep{kratter:2016}.
\citet{trapman:2017}, using additional constraints on the vertical structure
from \citet{kama:2016} and adding HD~2--1 line observations, suggest a gas mass,
in between these two extremes, of $(6\textendash 9)\times10^{-3}~\sun{}$. Recent
carbon sulfide (CS) molecular observations find a minimum disc mass of
$3\times10^{-4}~\sun{}$ \citep{teague:2018a}.

The characteristic timescale for aerodynamic drag to act on dust grains is
determined by the dimensionless stopping time, or Stokes number, $\st{}$
\citep{weidenschilling:1977, takeuchi:2002}. The Stokes number controls the rate
of vertical settling and radial drift. The Stokes number is proportional to the
grain size and inversely proportional to the gas density. Small grains ($\sim$
$\mu$m) experience high drag and have low $\st{}$. Whereas large grains
($\gtrsim$~cm) are largely decoupled from the gas phase and have high $\st{}$.
Grains with $\st{}\sim1$ experience the greatest rate of settling and drift.
In the presence of pressure bumps, $\st{}\sim1$ grains form axisymmetric
rings \citep{ayliffe:2012, dipierro:2015}.  The different response of small and
large grains to gas drag can be used to infer the mechanism for the origin of
the gaps.

To reproduce the axisymmetric gaps observed in recent ALMA observations, various
mechanisms have been proposed, including: planet-disc interactions
\citep{dipierro:2015}, self-induced dust trapping \citep{gonzalez:2017},
vortices \citep{zhu:2014}, condensation fronts \citep{zhang:2015}, non-ideal MHD
effects \citep{bethune:2016} and zonal flows \citep{johansen:2009,flock:2015}.

In this Letter, we explore the hypothesis that the axisymmetric rings and gaps
in the TW~Hya disc are carved by planets. A possible argument in favour of
planets is that the period ratio of the two inner planets is
${(41/24)}^{3/2}\approx2.2$ which is near the peak in distribution of period
ratios of \emph{Kepler} planet pairs \citep{winn:2015}. Our approach is similar
to \citet{dipierro:2015} who explored a similar hypothesis for HL~Tau. We aim to
constrain the planet masses required to explain the observational data on TW~Hya
and to motivate follow up observations.

\section{Methods}
\label{sec:methods}

\subsection{Numerical method}

We perform 3D global simulations of a dusty gas disc with embedded protoplanets
using \phantom{}, a smoothed particle hydrodynamics (SPH) code
\citep{price:2018a}. Dust interacts with the gas via a drag force. This allows
the dust to settle to the midplane and to migrate radially. We include back
reaction of dust on the gas, with the caveat that we model each grain size
independently. The dust also interacts gravitationally with the central star
and embedded planets. We use a low disc mass, so the disc is not
self-gravitating.

We simulate two dust grain sizes in separate calculations: 100~$\mu$m grains
with $\st{}\sim 0.3$ and 1~mm grains with $\st{}\sim 3$. We then combine the
results of the 100~$\mu$m and 1~mm calculations for radiative transfer
post-processing. These grains have Stokes number near unity to ensure efficient
settling and radial migration of our simulated grains. In this regime it is
appropriate to use the two-fluid method \citep{laibe:2012a}. We use $10^7$
particles for the gas, and $2.5\times 10^5$ for the dust. We use a greater
number of gas particles to prevent dust becoming trapped under the gas
resolution scale \citep{laibe:2012a}. For 100~$\mu$m- and 1~mm-sized grains, the
gas mean free path is large compared with the grain size, and so we assume
Epstein drag \citet{epstein:1924}. We assume spherical grains with a material
density of $3\ $g~cm$^{-3}$. We also perform gas-only simulations to explore the
impact of the outer planet.

We use sink particles \citep{bate:1995} to represent the central star and three
embedded protoplanets. The sink particles interact gravitationally with the disc
and with each other. Gravitationally bound dust and gas within the accretion
radius is accreted onto the sink. For computational efficiency, we set the
stellar accretion radius to be the inner edge of the disc.

\begin{figure}
   \begin{center}
      \includegraphics[height=0.460\columnwidth]{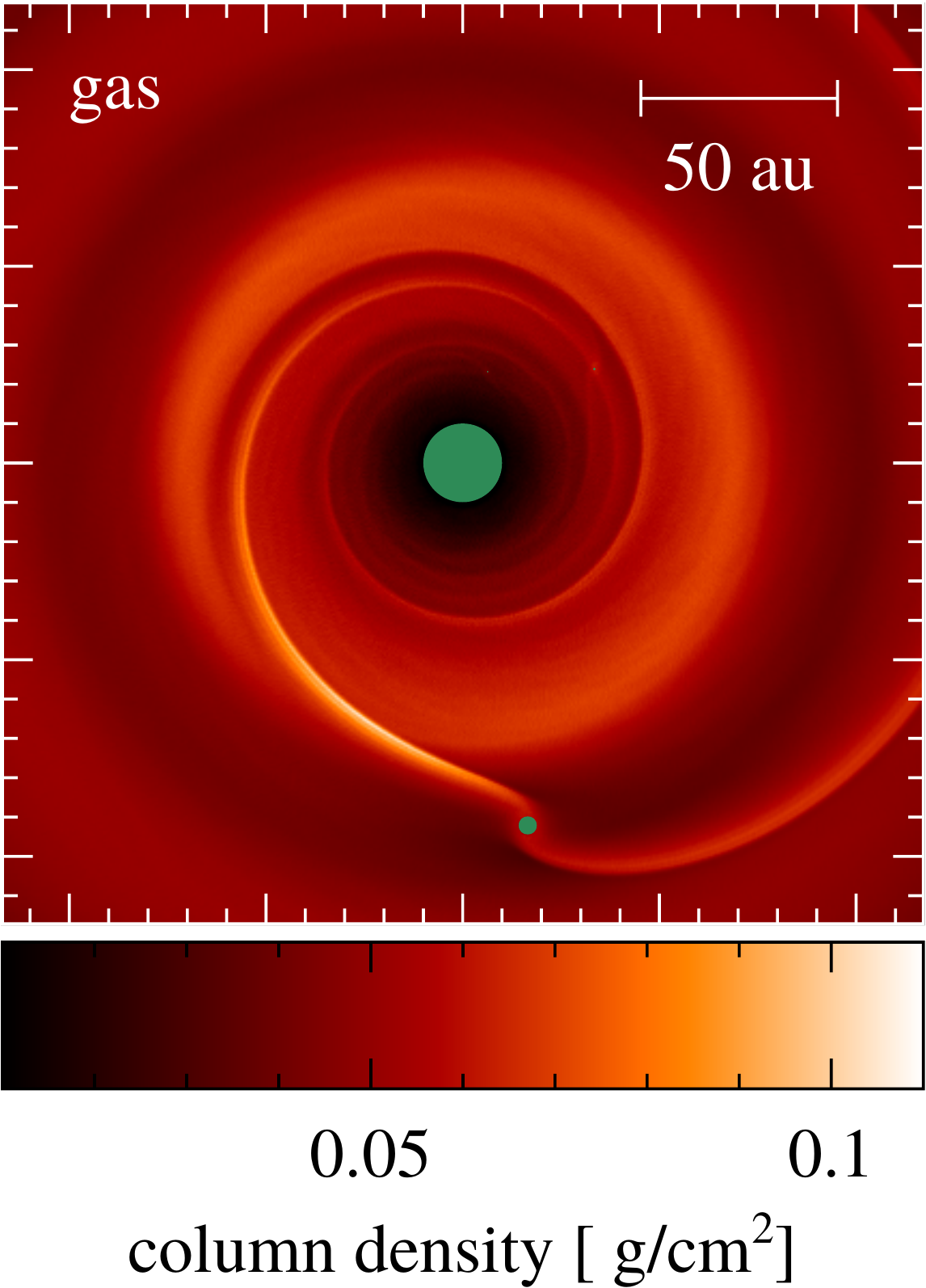}
      \includegraphics[height=0.460\columnwidth]{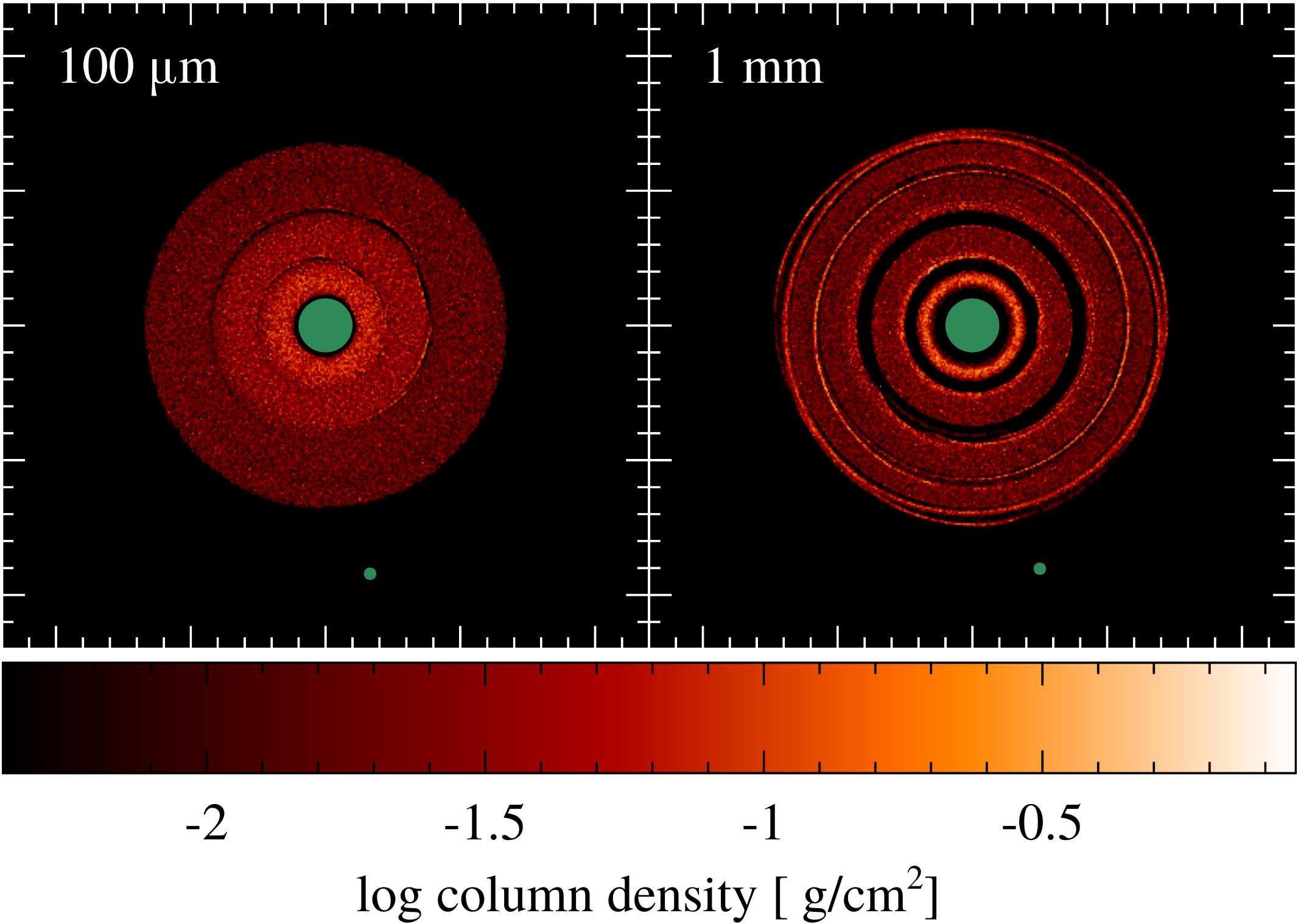}
      \caption{Gas (\textit{left}), 100~$\mu$m dust (\textit{center}), and 1~mm
         dust (\textit{right}) surface density for the model with 4~$\earth{}$
         inner planets (24 and 41~au) and 0.3~$\mathrm{M_J}$ outer planet
         (94~au) after $29400$~years. The green markers are sink particles with
         radius proportional to accretion radius. We do not model the inner
         ($\lesssim 10$~au) disc. The outer edge of the dust disc is $\sim
         70$~au.\label{fig:surface-density}}
   \end{center}
\end{figure}

\begin{figure*}
   \begin{center}
      \includegraphics[height=0.245\textwidth]{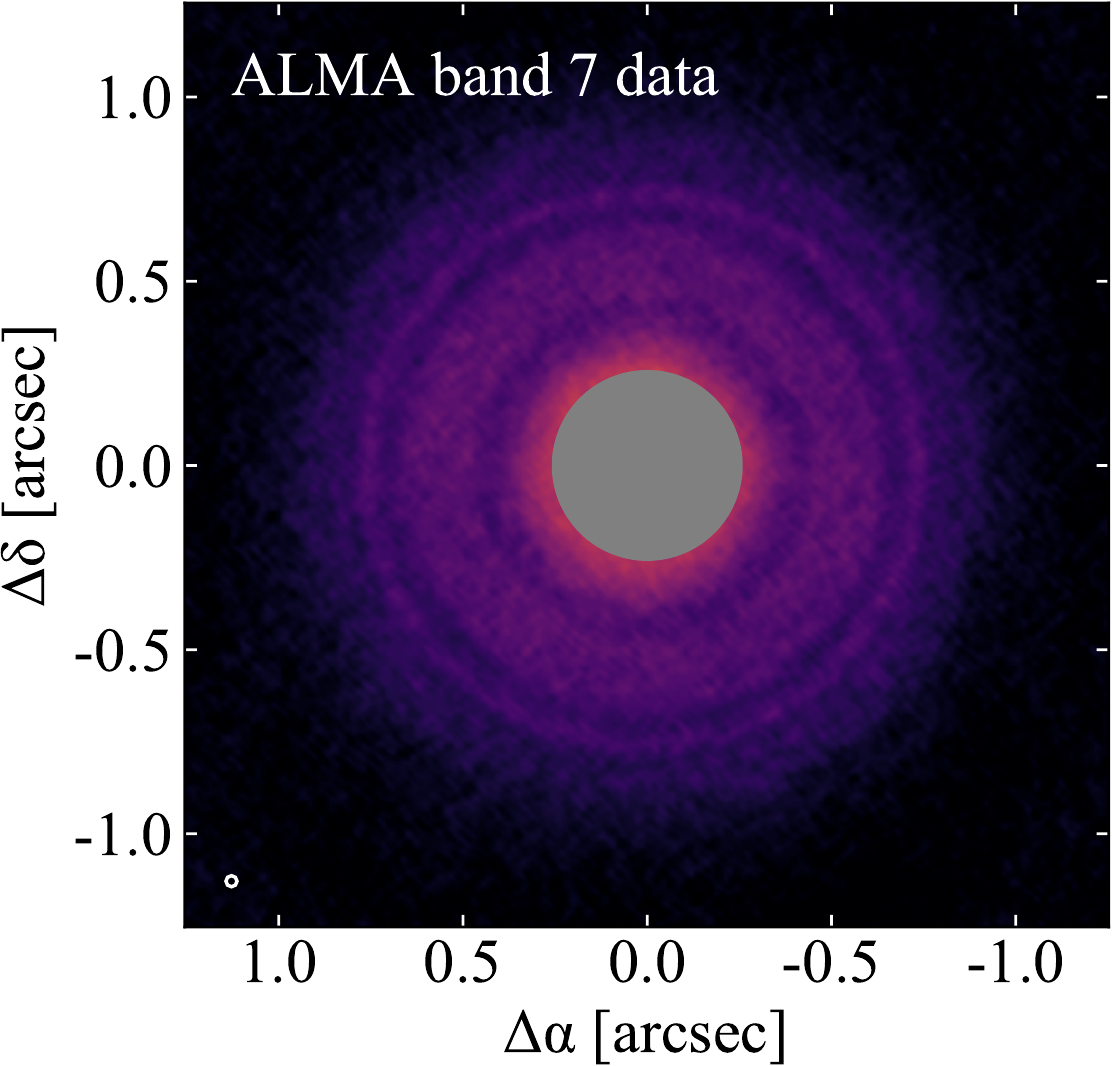} \quad
      \includegraphics[height=0.245\textwidth]{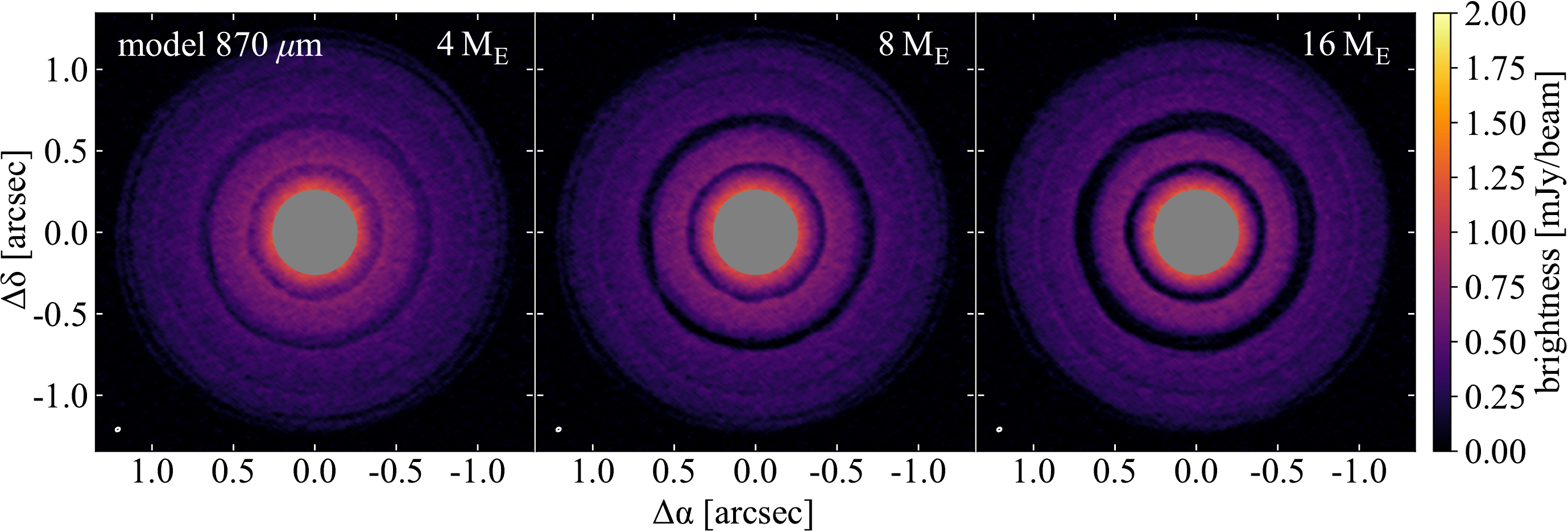}
      \caption{Inner planets (24 and 41~au). Synthetic observations of dust
         thermal emission at $870\,\mu$m, compared with ALMA band 7 observations
         from \citet{andrews:2016}. From left to right: dust+gas models with 4,
         8, 16~$\earth$ inner planets. The beam has FWHM 28$\times$21~mas in the
         model image, compared with 30~mas FWHM ($1.6$~au) circular beam in the
         observations. We obscured the inner $\approx$15~au as we did not model
         that region.\label{fig:alma}}
   \end{center}
\end{figure*}

\subsection{Initial conditions}

We assume a distance of 59.5~pc \citep{gaia-collaboration:2016} and a stellar
mass of $0.8\,\sun{}$ \citep{andrews:2012}. Scattered light and CO line
observations show that the gas disc extends out to at least $\sim$200~au
\citep{thi:2010} so we take the outer edge of the gas disc to be 200~au. We set
the inner edge of the disc to be $R_{\mathrm{in}}=10$~au for computational
efficiency. We do not attempt to model the inner disc ($\lesssim$~10~au) in this
study.

We set up a disc consisting of SPH particles following \citet{lodato:2010}. We
assume a gas mass of $7.5\times 10^{-4}\,\sun{}$ between 10~au and 200~au.
Extrapolating to 1~au implies a total disc mass higher by 1--10\%
depending on the surface density prescription, in the low end of the
\citet{thi:2010} range, but above the \citet{teague:2018a} minimum.  We set the
initial surface density profile as a smoothed power law: $\Sigma =
\Sigma_{\mathrm{in}} {(R/R_{\mathrm{in}})}^{-p} (1-\sqrt{R_{\mathrm{in}}/R})$,
where we adopt a shallow surface density profile with $p=0.5$. This, with our
gas disc mass, gives a surface density of $\Sigma \approx 0.05-0.08\
$g~cm${}^{-2}$, corresponding to a Stokes number of $\st{}\approx 0.25-0.4$ for
100~$\mu$m grains.

We assume a vertically isothermal equation of state $P={[c_s(R)]}^2\rho$ with $T
= 30\,\mathrm{K} {(R/R_{\mathrm{in}})}^{-0.25}$ where $c_s^2={k_B}T/\mu m_p$ and
$\mu=2.381$. This matches the CO snowline (20~K at 19~au) from
\citet{vant-hoff:2017} together with a midplane temperature of 15~K at 60~au
following previous modelling \citep{andrews:2012}. From these we infer a disc
aspect ratio of $H/R = c_s/(\Omega R) = 0.034$ at $R_{\mathrm{in}}$.
\citet{flaherty:2018} provide an upper limit on the turbulent velocity in the
outer disc of $v_{\mathrm{turb}}/c_s\approx 0.04-0.13$. This corresponds to an
$\alpha \sim {(v_{\mathrm{turb}}/c_s)}^2 \lesssim 0.002-0.02$. We choose a disc
viscosity \citep{shakura:1973} consistent with this upper limit and set the SPH
artificial viscosity to $\alpha_{\mathrm{AV}} = 0.1$ giving $\alpha \sim 0.001$.

The dust disc is more compact than the gas disc. Thermal dust emission shows
that the sub-mm dust disc extends to $\sim$ 50~au \citep{andrews:2016}. We set
the outer edge of the dust disc to $R_{\mathrm{out}}=80$~au, just inside the
orbital radius of the outer planet. This is to allow for some radial drift,
without having to follow the drift of dust particles from the gas outer radius.
We use the same inner edge as for the gas disc. Dust disc mass estimates are in
the range $(2$--$6)\times 10^{-4}\,\sun{}$ \citep{calvet:2002, thi:2010}. With
our gas disc mass this gives a dust-to-gas ratio of $\approx$~0.25--0.8 which is
one to two orders of magnitude higher than the typical interstellar value.
However, TW~Hya is an old disc within which we can expect significant evolution
away from its initial conditions. We set the dust-to-gas ratio (for 100~$\mu$m
and for 1~mm grains) to 0.05.

\subsection{Embedded planets}

We assume two super-Earth to super-Neptune mass planets at 24~au and 41~au,
respectively, to reproduce the two main observed gaps in sub-mm emission
\citep{andrews:2016}. We explored masses in the range of 4--16~$\earth{}$ for
these planets. To reproduce the outer gap observed in scattered light
\citep{van-boekel:2017} we placed a more massive planet at 94~au. We explored a
range of masses for the outer planet between 0.1--$2~\mathrm{M_J}$. We set the
planetary accretion radius $R_{\mathrm{acc}}$ to half the Hill radius,
$R_{\mathrm{H}} = a\sqrt[3]{m_{\mathrm{p}}/3M_*}$, where $a$ is the semi-major
axis, and $m_{\mathrm{p}}$ and $M_*$ are the planet mass and stellar mass,
respectively. Accretion proceeds unchecked for particles within 80\% of the
accretion radius.

\subsection{Synthetic observations}

We use the 3D radiative transfer code \mcfost{} \citep{pinte:2006, pinte:2009}
to post-process the \phantom{} output to produce simulated ALMA band 7 images,
CO maps and polarized scattered light images. We use a Voronoi (unstructured)
mesh using Voro++ \citep{rycroft:2009} in which the computational domain is
subdivided into cells generated from the positions of the SPH gas particles. We
assume an inclination of $5^{\circ}$ and position angle $152^{\circ}$
\citep{huang:2018} when making synthetic observations of the disc.

Within each cell we split the distribution of dust grain sizes into 100
logarithmic bins from 0.03~$\mu$m to 1~mm. Grains smaller than 1~$\mu$m are
assumed to trace the gas. Grains larger than 100~$\mu$m are interpolated between
100~$\mu$m and 1~mm simulations. Grains of intermediate size are interpolated
between gas and 100~$\mu$m dust. Total dust mass is set to $2.5\times
10^{-4}\,\sun{}$. We use $10^7$ photon packets to determine the temperature and
to produce synthetic observations.

We use the Common Astronomy Software Application (CASA) ALMA simulator (version
4.7) to produce synthetic band 7 ALMA images at 870~$\mu$m to compare with
\citet{andrews:2016}. We use a transit duration of 45 minutes, add thermal noise
from the receivers and atmosphere, and set the precipitable water vapour to
$0.5$~mm. To match the beam size of the observations, we choose an ALMA antenna
configuration (cycle 3.8) which gives a beam of FWHM 28$\times$21~mas at
$\mathrm{PA}=-60.3^{\circ}$.

We post-process gas-only \phantom{} simulations in \mcfost{} to produce
polarized scattered light images, and CO emission maps, assuming the dust
follows the gas. In these calculations we assume a dust-to-gas ratio of 0.01.
From 1.6$\,\mu$m (H-band) scattered light maps we calculated the azimuthal
Stokes component $Q_{\phi}$. We then add Gaussian noise, convolve with a
Gaussian beam with a FWHM of $48.5$~mas, and scale by $R^2$, following the
H-band SPHERE observations in \citet{van-boekel:2017}.

We also produce CO emission maps in the J~=~3--2 line. We assume
$T_{\mathrm{gas}} = T_{\mathrm{dust}}$ and that the emission is at LTE, as we
are looking at low-J CO lines. We assume a CO-to-H${}_2$ molecular abundance of
$10^{-4}$. We produce channel maps at 0.1~km/s resolution to then calculate the
$M_0$ moment map. We convolve with a Gaussian beam with FWHM of
139$\times$131~mas with a $\mathrm{PA}=-74.9^{\circ}$ following the ALMA
observations presented by \citet{huang:2018}.

\begin{figure*}
   \begin{center}
      \includegraphics[height=0.205\textwidth]{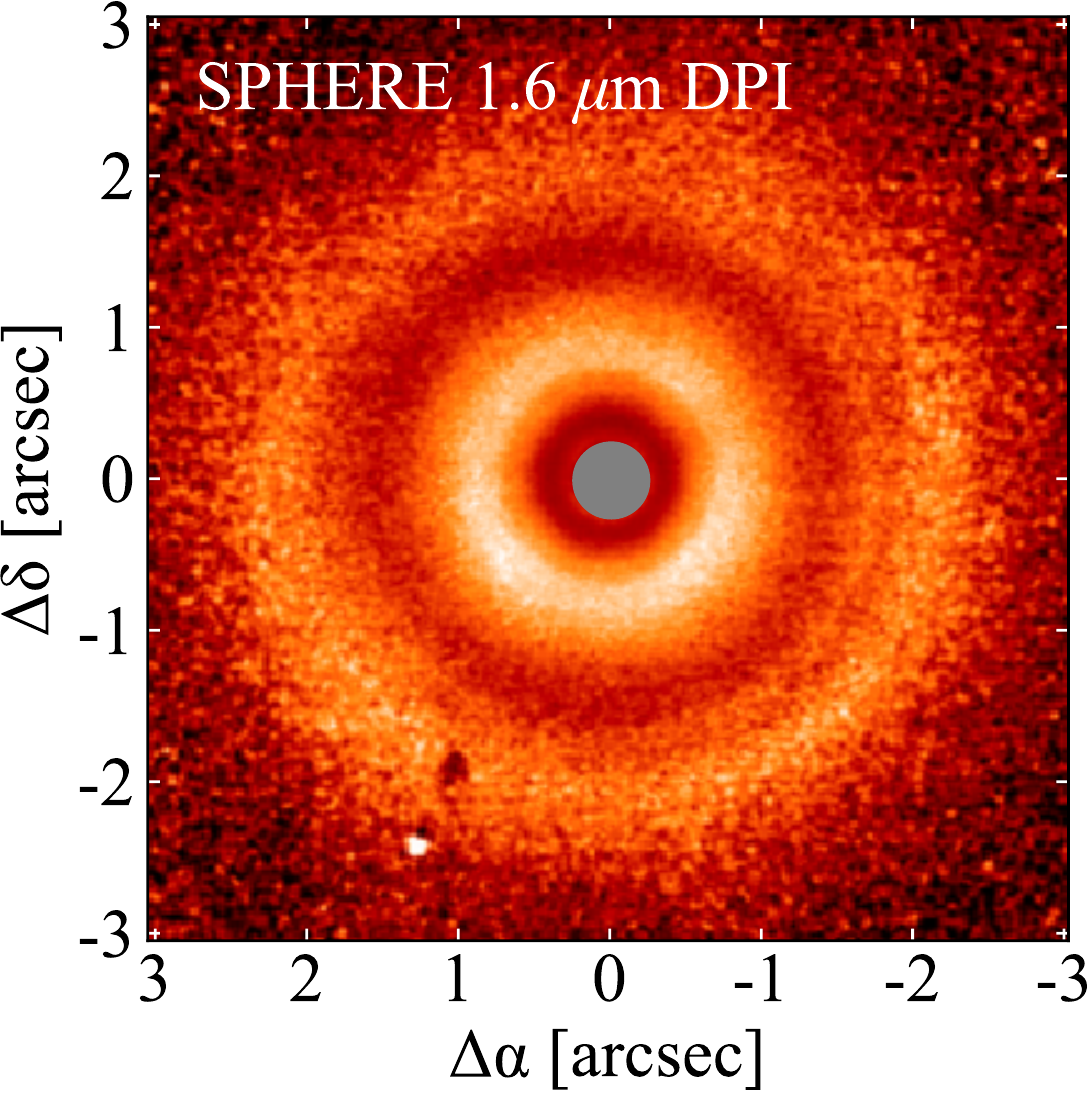} \quad
      \includegraphics[height=0.205\textwidth]{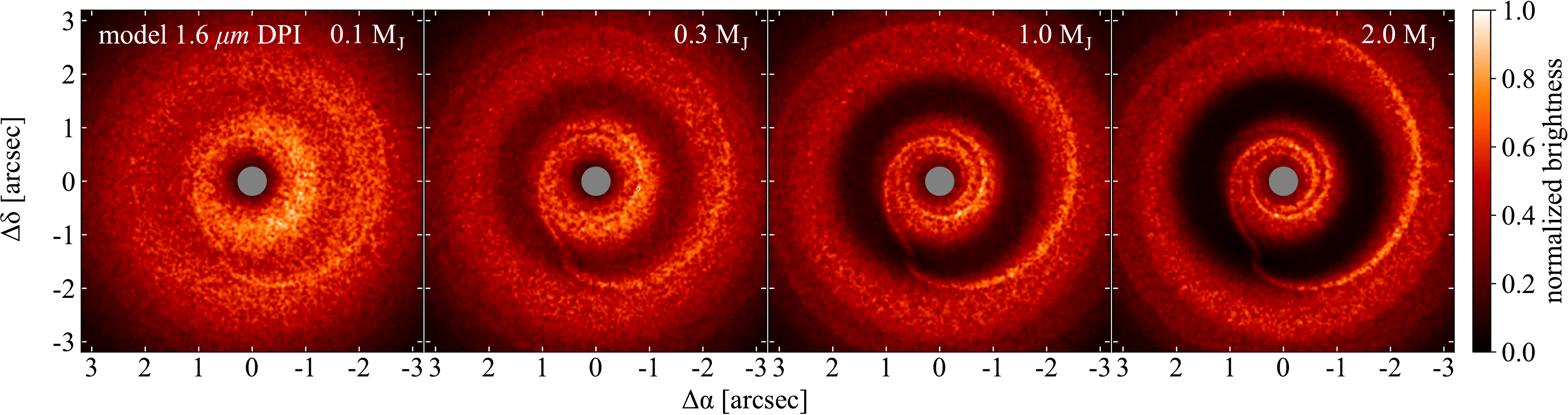}
      \includegraphics[height=0.203\textwidth]{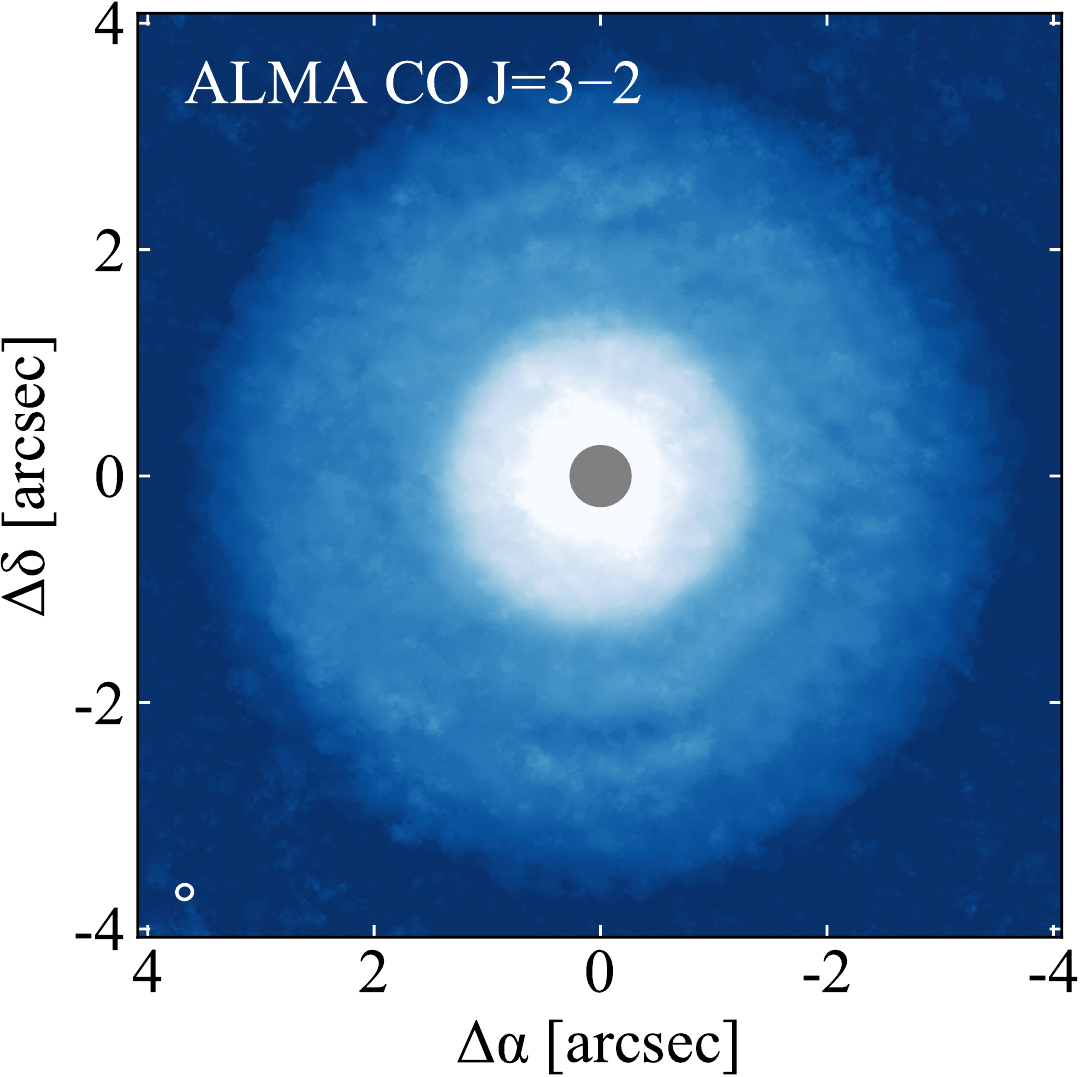} \quad
      \includegraphics[height=0.203\textwidth]{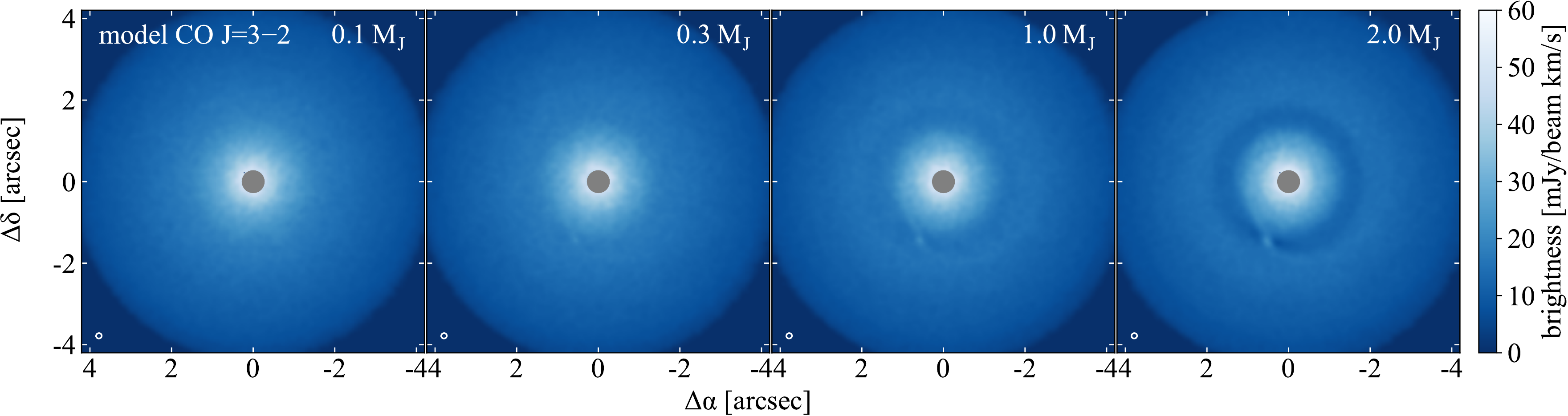}
      \caption{Gas-only models with outer planet (94~au) masses 0.1, 0.3, 1, and
         2~$\mathrm{M_J}$ after 100 orbits at 94~au. \textit{Top}: Comparison of
         synthetic observations of $1.6$~$\mu$m polarized intensity scaled by
         $R^2$ with the SPHERE observation. We convolved with a circular
         Gaussian beam of FWHM $48.5$~mas, and added noise. \textit{Bottom}:
         Comparison of synthetic CO J~=~3--2 integrated intensity emission maps
         with the ALMA observation. We convolved with a Gaussian beam of
         139$\times$131~mas with a $\mathrm{PA} = -74.9^{\circ}$. Top left panel
         is reproduced from \citet{van-boekel:2017} and bottom left is from data
         presented in \citet{huang:2018}.\label{fig:scattered-CO}}
   \end{center}
\end{figure*}

\section{Results}
\label{sec:results}

Figure~\ref{fig:surface-density} shows the gas and dust surface density after
29400~years (250, 100, and 32~orbits of the 24, 41, and 94~au planets, resp.)
for the model with 4~$\earth{}$ inner planets (24 and 41~au) and
0.3~$\mathrm{M_J}$ outer planet (94~au). The dust disc extends to $\sim 70$~au
(\textit{right} of Figure~\ref{fig:surface-density}). We observe cleared dust
gaps at the locations of the two inner planets, while the planets are not
massive enough to carve gaps in the gas (Figure~\ref{fig:surface-density}).
A possible caveat is that the gap profile of the innermost planet may be
affected by the inner boundary condition. The Saturn-to-Jupiter-mass outer
planet carves a (partial) gap in the gas, and produces a spiral density wave.
The region interior to 10~au is devoid of dust merely because it is within the
accretion radius of the stellar sink particle.

The inner planets (24 and 41~au) accreted $\approx$~10--20\% over the simulation
time. For models with an initial mass of 4, 8, and 16~$\earth{}$, the 24~au
planet accreted 0.4, 0.9, and 1.8~$\earth{}$, respectively, and the 41~au planet
accreted 1.0, 2.0, and 3.3~$\earth{}$, respectively. The outer planet (94~au)
accreted 65\%, 45\%, 20\%, and 10\% for models with initial mass 0.1, 0.3, 1,
and 2~$\mathrm{M_J}$, respectively. Planet migration was negligible.

\subsection{Dust thermal emission}

Figure~\ref{fig:alma} compares our synthetic band 7 ALMA observations of dust
thermal continuum emission for models with 4, 8, and 16~$\earth{}$ inner planets
with the ALMA observations \citep{andrews:2016}. Low mass planets ($<
0.1\,\mathrm{M_J} \approx 32\,\earth{}$) successfully reproduce the width and
axisymmetry of the gaps at 24 and 41~au.

Increasing the planet mass increases the gap width, as expected. Each planet
mass produces axisymmetric gaps. However, only the 4~$\earth{}$ planet produces
a partially cleared gap (like the ALMA observation). This is due to the fact
that, at 4~$\earth{}$, the planet is not large enough to carve a fully opened
gap in the 100~$\mu$m dust disc, but it is large enough to do so in the 1~mm
dust disc. The gap width at both 24 and 41~au is $\approx\,$5~au which is
consistent with the ALMA gap widths. However, the 41~au ALMA gap is narrower
than the 24~au gap, unlike our model, which suggests that the innermost planet
is the more massive of the two. This finding is consistent with the scattered
light observations, which show a low contrast gap at 24~au but none at 41~au. A
planet mass of 4~$\earth{}$ for the innermost planet is also consistent with an
upper limit suggested by \citet{nomura:2016}. However, it is not consistent with
modelling from \citet{van-boekel:2017} following \citet{duffell:2015}, and with
the low-viscosity models of \citet{dong:2017b}.

\subsection{Scattered light and CO emission}

\begin{figure}
   \begin{center}
      \includegraphics[width=1.00\columnwidth]{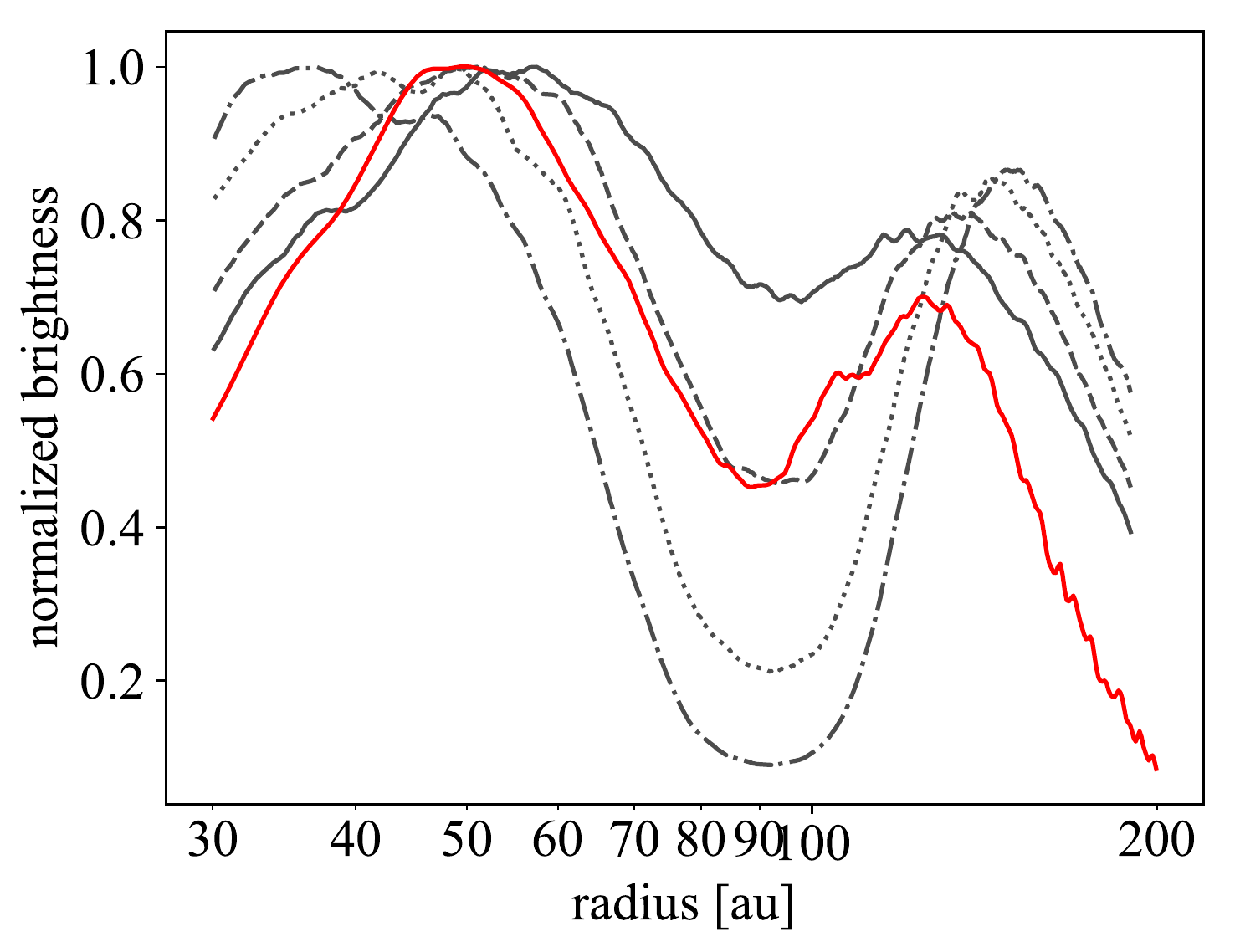}
      \caption{Azimuthally-averaged profiles of $1.6$~$\mu$m polarized intensity
         scaled by $R^2$, normalized to the peak at $\approx$~40--50~au. Solid,
         dashed, dotted, and dot-dashed lines are for 0.1, 0.3, 1, and
         2~$\mathrm{M_J}$ models after 100 orbits at 94~au. Red shows H-band
         data from \citet{van-boekel:2017}.\label{fig:scattered-radial}}
   \end{center}
\end{figure}

Figure~\ref{fig:scattered-CO} (\textit{top}) compares our synthetic polarized
scattered light H-band observations for gas-only models with outer planet masses
0.1, 0.3, 1, and 2~$\mathrm{M_J}$ after 102000~years, i.e. 100 orbits at 94~au,
with the SPHERE observation from \citet{van-boekel:2017}. The spiral arm induced
by the outer planet is visible in all our synthetic observations. For the 0.3,
1, and 2~$\mathrm{M_J}$ we observe a dip in scattered light at the orbital
radius of the planet. Figure~\ref{fig:scattered-radial} quantifies this by
comparing the azimuthally-averaged brightness profiles. The brightness contrast
between the peak and gap for a Saturn mass ($\approx$0.3~$\mathrm{M_J}$) planet
is consistent with the SPHERE observation.  A 0.1~$\mathrm{M_J}$ planet, by
contrast, fails to reproduce the gap.

Figure~\ref{fig:scattered-CO} (\textit{bottom}) compares synthetic CO~J~=~3--2
emission maps for gas-only models with outer planet masses 0.1, 0.3, 1, and
2~$\mathrm{M_J}$, with the ALMA observations \citep{huang:2018}. The
0.3~$\mathrm{M_J}$ model, which best fits the scattered light radial profile, is
consistent with CO observations. For a planet larger than 0.3~$\mathrm{M_J}$,
the gas surrounding the planet is visible in the $M_0$ map. This is because our
model has infinite signal-to-noise. Gas near the planet has perturbed velocity
and emits in a large number of channels. While the signal in each channel is
faint and would not be detected by ALMA, when aggregated in the $M_0$ map, it
becomes visible. Higher S/N ALMA observations might be able to detect the
planet.

\section{Discussion}
\label{sec:discussion}

For computational efficiency we did not model the inner disc (within
$\sim10$~au). This leads to a hotter temperature at radii $\lesssim 20$~au,
where the stellar radiation penetrates to the midplane. Thus the dust thermal
emission (Figure~\ref{fig:alma}) within the innermost planet orbit is
larger than the observation. At radii $>20$~au, we recover the vertically
stratified thermal structure expected for an optically thick disc. This
indicates that direct star light is not penetrating the midplane, and that the
temperature in this region of the disc does not depend anymore on the details of
the inner disc.

The overall flux is consistent to within a factor of 2 of the ALMA observation.
The contrast in flux between gaps and rings is $\approx$25\%, greater than the
observed contrast of 5--20$\%$ \citep{andrews:2016}. Our model has only two
grain sizes in the range that contribute emission in ALMA band 7.  Multigrain
dust simulations that include a greater range of grain sizes contributing to
emission, and calculate the collective back reaction of all dust grain
sizes on the gas, may alleviate that problem \citep{hutchison:2018}.

Spectral index observations from \citet{huang:2018} suggest that within the gaps
the maximum grain size is at most a few mm, whereas in the bright rings cm
grains are present. Therefore, the disc mass may be an order of magnitude higher
than our assumed mass, such that mm~grains have Stokes number corresponding to
that of 100~$\mu$m grains in our calculations. For a 4~$\earth{}$ planet, the
gaps would contain mm grains but no cm grains, as inferred from observations
\citep{huang:2018}.

\begin{figure}
   \begin{center}
      \includegraphics[width=1.00\columnwidth]{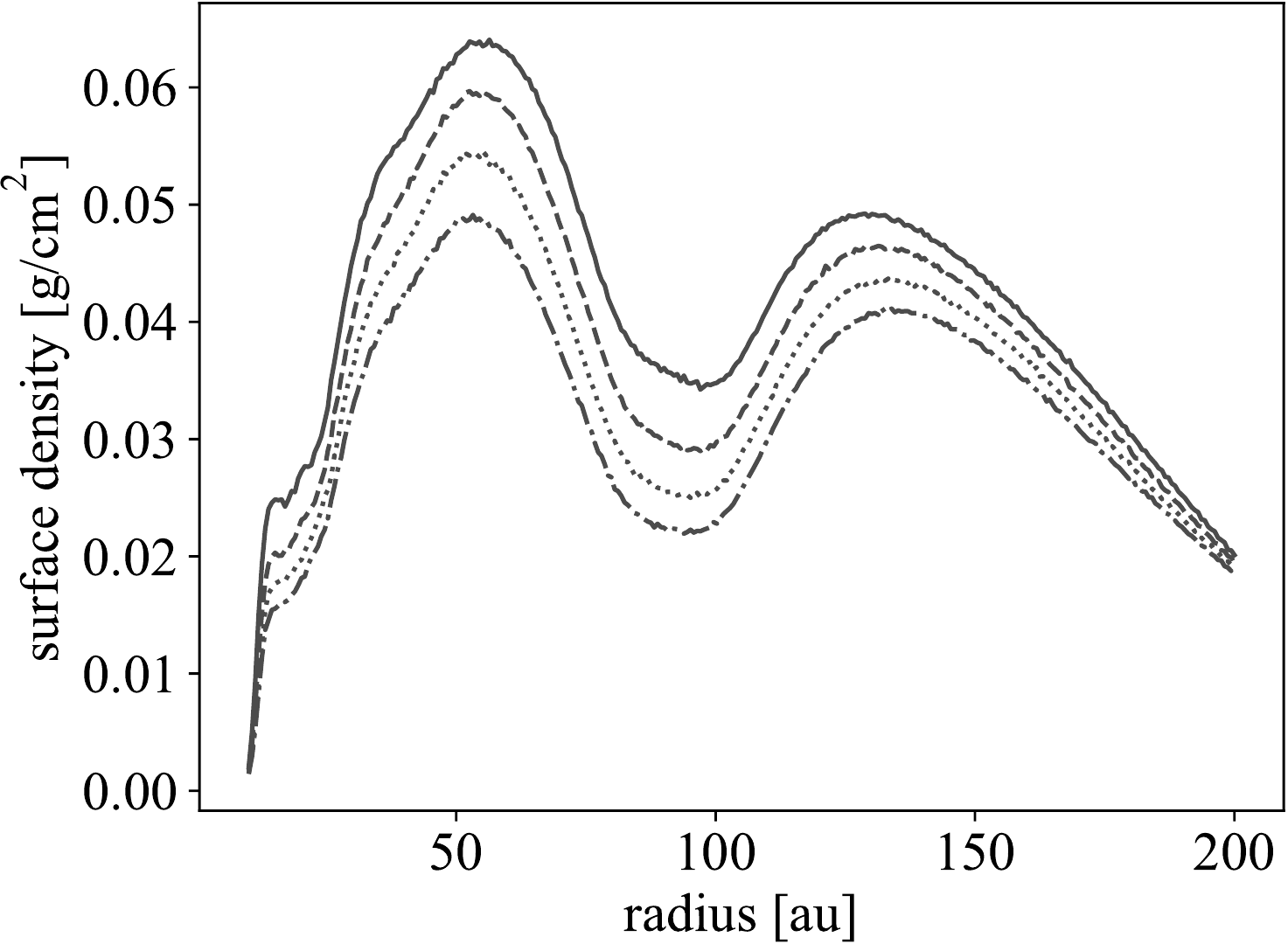}
      \caption{Surface density profile for gas-only model with
         0.3~$\mathrm{M_J}$ planet at 94~au. Top to bottom lines are at 40, 60,
         80, and 100 orbits at 94~au.\label{fig:gap}}
   \end{center}
\end{figure}

There is a tension between the outer planet mass required to reproduce the gap
in scattered light and CO observations, and the mass required to hide a spiral
arm. The synthetic observation from the $0.3~\mathrm{M_J}$ model (top of
Figure~\ref{fig:scattered-CO}) shows a greater degree of azimuthal asymmetry
than the SPHERE observation. Models with a lower mass planet ($\sim
0.1\,\mathrm{M_J}$) are more azimuthally symmetric. However, at those masses we
fail to reproduce the gap in both scattered light and in CO emission. A mass of
0.3~$\mathrm{M_J} \approx 95$~$\earth{}$ is higher than suggested by previous
authors \citep{dong:2017b, van-boekel:2017}. It is possible that our
calculations need to run for longer to reach a steady state gap profile.
Figure~\ref{fig:gap} shows the evolution of azimuthally-averaged surface
density. If the gap is not fully opened on the timescale of the simulation then
we can only put an upper limit on the planet mass. It is also possible that we
are overestimating the planet mass if the gap were accentuated by shadowing from
the inner disc \citep{debes:2013, debes:2017, poteet:2018}.  The pebble
isolation mass \citep{bitsch:2018} for 100~$\mu$m and 1~mm grains is
$\approx$20~$\earth{}$ which is well below each model presented here.  This
suggests that inward radial drift of these grains occurred before the planet
reached its current mass.

For our disc model, the stellar accretion rate is $1.5\times
10^{-10}\,\sun{}\,\mathrm{yr}^{-1}$ which is an order of magnitude below the
estimated rate \citep{brickhouse:2012}. The accretion rate is given by $\dot{M}
= 3\pi\Sigma\alpha c_s H$. This suggests two modifications to increase
$\dot{M}$: we could increase the disc mass, and we could increase the disc
viscosity. As discussed, a ten-fold increase in disc mass is possible, given
spectral index observations. Line width observations provide an upper limit
to the disc viscosity \citep{flaherty:2018}. An alternative may be that
accretion is driven by winds \citep{simon:2018}.

Increasing the stellar accretion rate via either approach increases the
planetary accretion rate. For the 4$\,\earth{}$ model the inner planets (24 and
41~au) accrete $0.4~\earth{}$ and $1.0~\earth{}$, respectively, over the
$\approx30000$~yrs of simulation time. Extrapolating this rate to a million
years leads to accretion of $\sim$~10--30~$\earth{}$, which is uncomfortably
high. Increasing the disc viscosity also requires larger planets to form gaps
initially as a greater gravitational torque is required to overcome the viscous
torque from the gas \citep{dipierro:2016}.

The product of planetary mass and accretion rate $M_p\dot{M_p}$ for the outer
planet (94~au) in the $0.3~\mathrm{M_J}$ model is $2\times 10^{-7}\,
\mathrm{M_J^2/yr}$, which is a factor of 5 greater than the upper limit deduced
from Keck/NIRC2 vortex coronagraph observations \citep{ruane:2017}. Given that
our model constrains the planet mass via the gap depth this suggests that the
accretion rate may be too high in our model. We use a relatively large sink
radius for computational reasons. A smaller sink radius may reduce the accretion
rate, and improve agreement with the observed value.

\section{Summary}
\label{sec:summary}

We have performed global three-dimensional SPH simulations of a dusty disc with
embedded protoplanets and produced synthetic observations of dust continuum, CO
emission, and polarized scattered light to test our model against recent
observations.

\begin{enumerate}
   \item We reproduce the gaps in dust emission in the ALMA observations of
      TW~Hya with two $\approx$4~$\earth{}$ planets at 24~au and 41~au.
   \item We show that a giant planet (0.1--0.3~$\mathrm{M_J}$) at 94~au can
      explain the main gap in scattered light observations, and is consistent
      with CO observations. However, a spiral arm is also evident, for which
      there is only tentative evidence in the SPHERE image.
   \item Our model requires a disc mass $\lesssim10^{-2}\,\sun{}$ in agreement
      with CO observations rather than the $>0.05\,\sun{}$ disc mass inferred by
      \citet{bergin:2013}. A low mass disc is consistent with recent constraints
      on disc turbulence \citep{flaherty:2018}.
\end{enumerate}

\section*{Acknowledgements}

DM is funded by a Research Training Program Stipend from the Australian
government. We acknowledge Australia Research Council funding via DP180104235,
FT130100034, and FT170100040. We used gSTAR and OzStar, funded by Swinburne
University of Technology and the Australian government, the MonARCH cluster at
Monash, and \textsc{splash} \citep{price:2007}.

\bibliography{references}

\begin{thebibliography}{}
\makeatletter
\relax
\def\mn@urlcharsother{\let\do\@makeother \do\$\do\&\do\#\do\^\do\_\do\%\do\~}
\def\mn@doi{\begingroup\mn@urlcharsother \@ifnextchar [ {\mn@doi@}
  {\mn@doi@[]}}
\def\mn@doi@[#1]#2{\def\@tempa{#1}\ifx\@tempa\@empty \href
  {http://dx.doi.org/#2} {doi:#2}\else \href {http://dx.doi.org/#2} {#1}\fi
  \endgroup}
\def\mn@eprint#1#2{\mn@eprint@#1:#2::\@nil}
\def\mn@eprint@arXiv#1{\href {http://arxiv.org/abs/#1} {{\tt arXiv:#1}}}
\def\mn@eprint@dblp#1{\href {http://dblp.uni-trier.de/rec/bibtex/#1.xml}
  {dblp:#1}}
\def\mn@eprint@#1:#2:#3:#4\@nil{\def\@tempa {#1}\def\@tempb {#2}\def\@tempc
  {#3}\ifx \@tempc \@empty \let \@tempc \@tempb \let \@tempb \@tempa \fi \ifx
  \@tempb \@empty \def\@tempb {arXiv}\fi \@ifundefined
  {mn@eprint@\@tempb}{\@tempb:\@tempc}{\expandafter \expandafter \csname
  mn@eprint@\@tempb\endcsname \expandafter{\@tempc}}}

\bibitem[\protect\citeauthoryear{{Andrews} et~al.,}{{Andrews}
  et~al.}{2012}]{andrews:2012}
{Andrews} S.~M.,  et~al., 2012, \mn@doi [\apj] {10.1088/0004-637X/744/2/162},
  \href {http://adsabs.harvard.edu/abs/2012ApJ...744..162A} {744, 162}

\bibitem[\protect\citeauthoryear{{Andrews} et~al.,}{{Andrews}
  et~al.}{2016}]{andrews:2016}
{Andrews} S.~M.,  et~al., 2016, \mn@doi [\apjl] {10.3847/2041-8205/820/2/L40},
  \href {http://adsabs.harvard.edu/abs/2016ApJ...820L..40A} {820, L40}

\bibitem[\protect\citeauthoryear{{Ayliffe}, {Laibe}, {Price}  \&
  {Bate}}{{Ayliffe} et~al.}{2012}]{ayliffe:2012}
{Ayliffe} B.~A.,  {Laibe} G.,  {Price} D.~J.,   {Bate} M.~R.,  2012, \mn@doi
  [\mnras] {10.1111/j.1365-2966.2012.20967.x}, \href
  {http://adsabs.harvard.edu/abs/2012MNRAS.423.1450A} {423, 1450}

\bibitem[\protect\citeauthoryear{{Barrado Y Navascu{\'e}s}}{{Barrado Y
  Navascu{\'e}s}}{2006}]{barrado-y-navascues:2006}
{Barrado Y Navascu{\'e}s} D.,  2006, \mn@doi [\aap]
  {10.1051/0004-6361:20065717}, \href
  {http://adsabs.harvard.edu/abs/2006A%26A...459..511B} {459, 511}

\bibitem[\protect\citeauthoryear{{Bate}, {Bonnell}  \& {Price}}{{Bate}
  et~al.}{1995}]{bate:1995}
{Bate} M.~R.,  {Bonnell} I.~A.,   {Price} N.~M.,  1995, \mn@doi [\mnras]
  {10.1093/mnras/277.2.362}, \href
  {http://adsabs.harvard.edu/abs/1995MNRAS.277..362B} {277, 362}

\bibitem[\protect\citeauthoryear{{Bergin} et~al.,}{{Bergin}
  et~al.}{2013}]{bergin:2013}
{Bergin} E.~A.,  et~al., 2013, \mn@doi [\nat] {10.1038/nature11805}, \href
  {http://adsabs.harvard.edu/abs/2013Natur.493..644B} {493, 644}

\bibitem[\protect\citeauthoryear{{B{\'e}thune}, {Lesur}  \&
  {Ferreira}}{{B{\'e}thune} et~al.}{2016}]{bethune:2016}
{B{\'e}thune} W.,  {Lesur} G.,   {Ferreira} J.,  2016, \mn@doi [\aap]
  {10.1051/0004-6361/201527874}, \href
  {http://adsabs.harvard.edu/abs/2016A%26A...589A..87B} {589, A87}

\bibitem[\protect\citeauthoryear{{Bitsch} et~al.,}{{Bitsch}
  et~al.}{2018}]{bitsch:2018}
{Bitsch} B.,  et~al., 2018, \mn@doi [\aap] {10.1051/0004-6361/201731931}, \href
  {http://adsabs.harvard.edu/abs/2018A%26A...612A..30B} {612, A30}

\bibitem[\protect\citeauthoryear{{Brickhouse} et~al.,}{{Brickhouse}
  et~al.}{2012}]{brickhouse:2012}
{Brickhouse} N.~S.,  et~al., 2012, \mn@doi [\apjl]
  {10.1088/2041-8205/760/2/L21}, \href
  {http://adsabs.harvard.edu/abs/2012ApJ...760L..21B} {760, L21}

\bibitem[\protect\citeauthoryear{{Calvet} et~al.,}{{Calvet}
  et~al.}{2002}]{calvet:2002}
{Calvet} N.,  et~al., 2002, \mn@doi [\apj] {10.1086/339061}, \href
  {http://adsabs.harvard.edu/abs/2002ApJ...568.1008C} {568, 1008}

\bibitem[\protect\citeauthoryear{{Debes} et~al.,}{{Debes}
  et~al.}{2013}]{debes:2013}
{Debes} J.~H.,  et~al., 2013, \mn@doi [\apj] {10.1088/0004-637X/771/1/45},
  \href {http://adsabs.harvard.edu/abs/2013ApJ...771...45D} {771, 45}

\bibitem[\protect\citeauthoryear{{Debes} et~al.,}{{Debes}
  et~al.}{2017}]{debes:2017}
{Debes} J.~H.,  et~al., 2017, \mn@doi [\apj] {10.3847/1538-4357/835/2/205},
  \href {http://adsabs.harvard.edu/abs/2017ApJ...835..205D} {835, 205}

\bibitem[\protect\citeauthoryear{{Dipierro} et~al.,}{{Dipierro}
  et~al.}{2015}]{dipierro:2015}
{Dipierro} G.,  et~al., 2015, \mn@doi [\mnras] {10.1093/mnrasl/slv105}, \href
  {http://adsabs.harvard.edu/abs/2015MNRAS.453L..73D} {453, L73}

\bibitem[\protect\citeauthoryear{{Dipierro}, {Laibe}, {Price}  \&
  {Lodato}}{{Dipierro} et~al.}{2016}]{dipierro:2016}
{Dipierro} G.,  {Laibe} G.,  {Price} D.~J.,   {Lodato} G.,  2016, \mn@doi
  [\mnras] {10.1093/mnrasl/slw032}, \href
  {http://adsabs.harvard.edu/abs/2016MNRAS.459L...1D} {459, L1}

\bibitem[\protect\citeauthoryear{{Dong} \& {Fung}}{{Dong} \&
  {Fung}}{2017}]{dong:2017b}
{Dong} R.,  {Fung} J.,  2017, \mn@doi [\apj] {10.3847/1538-4357/835/2/146},
  \href {http://adsabs.harvard.edu/abs/2017ApJ...835..146D} {835, 146}

\bibitem[\protect\citeauthoryear{{Duffell} \& {Dong}}{{Duffell} \&
  {Dong}}{2015}]{duffell:2015}
{Duffell} P.~C.,  {Dong} R.,  2015, \mn@doi [\apj]
  {10.1088/0004-637X/802/1/42}, \href
  {http://adsabs.harvard.edu/abs/2015ApJ...802...42D} {802, 42}

\bibitem[\protect\citeauthoryear{{Epstein}}{{Epstein}}{1924}]{epstein:1924}
{Epstein} P.~S.,  1924, \mn@doi [Physical Review] {10.1103/PhysRev.23.710},
  \href {http://adsabs.harvard.edu/abs/1924PhRv...23..710E} {23, 710}

\bibitem[\protect\citeauthoryear{{Flaherty} et~al.,}{{Flaherty}
  et~al.}{2018}]{flaherty:2018}
{Flaherty} K.~M.,  et~al., 2018, \mn@doi [\apj] {10.3847/1538-4357/aab615},
  \href {http://adsabs.harvard.edu/abs/2018ApJ...856..117F} {856, 117}

\bibitem[\protect\citeauthoryear{{Flock} et~al.,}{{Flock}
  et~al.}{2015}]{flock:2015}
{Flock} M.,  et~al., 2015, \mn@doi [\aap] {10.1051/0004-6361/201424693}, \href
  {http://adsabs.harvard.edu/abs/2015A%26A...574A..68F} {574, A68}

\bibitem[\protect\citeauthoryear{{Gaia Collaboration} et~al.,}{{Gaia
  Collaboration} et~al.}{2016}]{gaia-collaboration:2016}
{Gaia Collaboration} et~al., 2016, \mn@doi [\aap]
  {10.1051/0004-6361/201629272}, \href
  {http://adsabs.harvard.edu/abs/2016A%26A...595A...1G} {595, A1}

\bibitem[\protect\citeauthoryear{{Gaia Collaboration} et~al.,}{{Gaia
  Collaboration} et~al.}{2018}]{gaia-collaboration:2018}
{Gaia Collaboration} et~al., 2018, \mn@doi [\aap]
  {10.1051/0004-6361/201833051}, \href
  {http://adsabs.harvard.edu/abs/2018A%26A...616A...1G} {616, A1}

\bibitem[\protect\citeauthoryear{{Gonzalez}, {Laibe}  \& {Maddison}}{{Gonzalez}
  et~al.}{2017}]{gonzalez:2017}
{Gonzalez} J.-F.,  {Laibe} G.,   {Maddison} S.~T.,  2017, \mn@doi [\mnras]
  {10.1093/mnras/stx016}, \href
  {http://adsabs.harvard.edu/abs/2017MNRAS.tmp...29G} {}

\bibitem[\protect\citeauthoryear{{Haisch}, {Lada}  \& {Lada}}{{Haisch}
  et~al.}{2001}]{haisch:2001}
{Haisch} Jr. K.~E.,  {Lada} E.~A.,   {Lada} C.~J.,  2001, \mn@doi [\apjl]
  {10.1086/320685}, \href {http://adsabs.harvard.edu/abs/2001ApJ...553L.153H}
  {553, L153}

\bibitem[\protect\citeauthoryear{{Huang} et~al.,}{{Huang}
  et~al.}{2018}]{huang:2018}
{Huang} J.,  et~al., 2018, \mn@doi [\apj] {10.3847/1538-4357/aaa1e7}, \href
  {http://adsabs.harvard.edu/abs/2018ApJ...852..122H} {852, 122}

\bibitem[\protect\citeauthoryear{{Hutchison}, {Price}  \& {Laibe}}{{Hutchison}
  et~al.}{2018}]{hutchison:2018}
{Hutchison} M.,  {Price} D.~J.,   {Laibe} G.,  2018, \mn@doi [\mnras]
  {10.1093/mnras/sty367}, \href
  {http://adsabs.harvard.edu/abs/2018MNRAS.tmp..359H} {}

\bibitem[\protect\citeauthoryear{{Johansen}, {Youdin}  \& {Klahr}}{{Johansen}
  et~al.}{2009}]{johansen:2009}
{Johansen} A.,  {Youdin} A.,   {Klahr} H.,  2009, \mn@doi [\apj]
  {10.1088/0004-637X/697/2/1269}, \href
  {http://adsabs.harvard.edu/abs/2009ApJ...697.1269J} {697, 1269}

\bibitem[\protect\citeauthoryear{{Kama} et~al.,}{{Kama}
  et~al.}{2016}]{kama:2016}
{Kama} M.,  et~al., 2016, \mn@doi [\aap] {10.1051/0004-6361/201526791}, \href
  {http://adsabs.harvard.edu/abs/2016A%26A...588A.108K} {588, A108}

\bibitem[\protect\citeauthoryear{{Kratter} \& {Lodato}}{{Kratter} \&
  {Lodato}}{2016}]{kratter:2016}
{Kratter} K.,  {Lodato} G.,  2016, \mn@doi [\araa]
  {10.1146/annurev-astro-081915-023307}, \href
  {http://adsabs.harvard.edu/abs/2016ARA%26A..54..271K} {54, 271}

\bibitem[\protect\citeauthoryear{{Laibe} \& {Price}}{{Laibe} \&
  {Price}}{2012}]{laibe:2012a}
{Laibe} G.,  {Price} D.~J.,  2012, \mn@doi [\mnras]
  {10.1111/j.1365-2966.2011.20202.x}, \href
  {http://adsabs.harvard.edu/abs/2012MNRAS.420.2345L} {420, 2345}

\bibitem[\protect\citeauthoryear{{Lodato} \& {Price}}{{Lodato} \&
  {Price}}{2010}]{lodato:2010}
{Lodato} G.,  {Price} D.~J.,  2010, \mn@doi [\mnras]
  {10.1111/j.1365-2966.2010.16526.x}, \href
  {http://adsabs.harvard.edu/abs/2010MNRAS.405.1212L} {405, 1212}

\bibitem[\protect\citeauthoryear{{Nomura} et~al.,}{{Nomura}
  et~al.}{2016}]{nomura:2016}
{Nomura} H.,  et~al., 2016, \mn@doi [\apjl] {10.3847/2041-8205/819/1/L7}, \href
  {http://adsabs.harvard.edu/abs/2016ApJ...819L...7N} {819, L7}

\bibitem[\protect\citeauthoryear{{Pinte}, {M{\'e}nard}, {Duch{\^e}ne}  \&
  {Bastien}}{{Pinte} et~al.}{2006}]{pinte:2006}
{Pinte} C.,  {M{\'e}nard} F.,  {Duch{\^e}ne} G.,   {Bastien} P.,  2006, \mn@doi
  [\aap] {10.1051/0004-6361:20053275}, \href
  {http://adsabs.harvard.edu/abs/2006A%26A...459..797P} {459, 797}

\bibitem[\protect\citeauthoryear{{Pinte} et~al.,}{{Pinte}
  et~al.}{2009}]{pinte:2009}
{Pinte} C.,  et~al., 2009, \mn@doi [\aap] {10.1051/0004-6361/200811555}, \href
  {http://adsabs.harvard.edu/abs/2009A%26A...498..967P} {498, 967}

\bibitem[\protect\citeauthoryear{{Poteet} et~al.,}{{Poteet}
  et~al.}{2018}]{poteet:2018}
{Poteet} C.~A.,  et~al., 2018, \mn@doi [\apj] {10.3847/1538-4357/aac2e4}, \href
  {http://adsabs.harvard.edu/abs/2018ApJ...860..115P} {860, 115}

\bibitem[\protect\citeauthoryear{{Price}}{{Price}}{2007}]{price:2007}
{Price} D.~J.,  2007, \mn@doi [\pasa] {10.1071/AS07022}, \href
  {http://adsabs.harvard.edu/abs/2007PASA...24..159P} {24, 159}

\bibitem[\protect\citeauthoryear{{Price} et~al.,}{{Price}
  et~al.}{2018}]{price:2018a}
{Price} D.~J.,  et~al., 2018, \mn@doi [\pasa] {10.1017/pasa.2018.25}, \href
  {http://adsabs.harvard.edu/abs/2018PASA...35...31P} {35, e031}

\bibitem[\protect\citeauthoryear{{Ruane} et~al.,}{{Ruane}
  et~al.}{2017}]{ruane:2017}
{Ruane} G.,  et~al., 2017, \mn@doi [\aj] {10.3847/1538-3881/aa7b81}, \href
  {http://adsabs.harvard.edu/abs/2017AJ....154...73R} {154, 73}

\bibitem[\protect\citeauthoryear{{Rycroft}}{{Rycroft}}{2009}]{rycroft:2009}
{Rycroft} C.~H.,  2009, \mn@doi [Chaos] {10.1063/1.3215722}, \href
  {http://adsabs.harvard.edu/abs/2009Chaos..19d1111R} {19, 041111}

\bibitem[\protect\citeauthoryear{{Shakura} \& {Sunyaev}}{{Shakura} \&
  {Sunyaev}}{1973}]{shakura:1973}
{Shakura} N.~I.,  {Sunyaev} R.~A.,  1973, \aap, \href
  {http://adsabs.harvard.edu/abs/1973A%26A....24..337S} {24, 337}

\bibitem[\protect\citeauthoryear{{Simon}, {Bai}, {Flaherty}  \&
  {Hughes}}{{Simon} et~al.}{2018}]{simon:2018}
{Simon} J.~B.,  {Bai} X.-N.,  {Flaherty} K.~M.,   {Hughes} A.~M.,  2018,
  \mn@doi [\apj] {10.3847/1538-4357/aad86d}, \href
  {http://adsabs.harvard.edu/abs/2018ApJ...865...10S} {865, 10}

\bibitem[\protect\citeauthoryear{{Takeuchi} \& {Lin}}{{Takeuchi} \&
  {Lin}}{2002}]{takeuchi:2002}
{Takeuchi} T.,  {Lin} D.~N.~C.,  2002, \mn@doi [\apj] {10.1086/344437}, \href
  {http://adsabs.harvard.edu/abs/2002ApJ...581.1344T} {581, 1344}

\bibitem[\protect\citeauthoryear{{Teague} et~al.,}{{Teague}
  et~al.}{2018}]{teague:2018a}
{Teague} R.,  et~al., 2018, \mn@doi [\apj] {10.3847/1538-4357/aad80e}, \href
  {http://adsabs.harvard.edu/abs/2018ApJ...864..133T} {864, 133}

\bibitem[\protect\citeauthoryear{{Thi} et~al.,}{{Thi} et~al.}{2010}]{thi:2010}
{Thi} W.-F.,  et~al., 2010, \mn@doi [\aap] {10.1051/0004-6361/201014578}, \href
  {http://adsabs.harvard.edu/abs/2010A%26A...518L.125T} {518, L125}

\bibitem[\protect\citeauthoryear{{Trapman} et~al.,}{{Trapman}
  et~al.}{2017}]{trapman:2017}
{Trapman} L.,  et~al., 2017, \mn@doi [\aap] {10.1051/0004-6361/201630308},
  \href {http://adsabs.harvard.edu/abs/2017A%26A...605A..69T} {605, A69}

\bibitem[\protect\citeauthoryear{{Weidenschilling}}{{Weidenschilling}}{1977}]{weidenschilling:1977}
{Weidenschilling} S.~J.,  1977, \mn@doi [\mnras] {10.1093/mnras/180.1.57},
  \href {http://adsabs.harvard.edu/abs/1977MNRAS.180...57W} {180, 57}

\bibitem[\protect\citeauthoryear{{Winn} \& {Fabrycky}}{{Winn} \&
  {Fabrycky}}{2015}]{winn:2015}
{Winn} J.~N.,  {Fabrycky} D.~C.,  2015, \mn@doi [\araa]
  {10.1146/annurev-astro-082214-122246}, \href
  {http://adsabs.harvard.edu/abs/2015ARA%26A..53..409W} {53, 409}

\bibitem[\protect\citeauthoryear{{Zhang}, {Blake}  \& {Bergin}}{{Zhang}
  et~al.}{2015}]{zhang:2015}
{Zhang} K.,  {Blake} G.~A.,   {Bergin} E.~A.,  2015, \mn@doi [\apjl]
  {10.1088/2041-8205/806/1/L7}, \href
  {http://adsabs.harvard.edu/abs/2015ApJ...806L...7Z} {806, L7}

\bibitem[\protect\citeauthoryear{{Zhu} \& {Stone}}{{Zhu} \&
  {Stone}}{2014}]{zhu:2014}
{Zhu} Z.,  {Stone} J.~M.,  2014, \mn@doi [\apj] {10.1088/0004-637X/795/1/53},
  \href {http://adsabs.harvard.edu/abs/2014ApJ...795...53Z} {795, 53}

\bibitem[\protect\citeauthoryear{{van Boekel} et~al.,}{{van Boekel}
  et~al.}{2017}]{van-boekel:2017}
{van Boekel} R.,  et~al., 2017, \mn@doi [\apj] {10.3847/1538-4357/aa5d68},
  \href {http://adsabs.harvard.edu/abs/2017ApJ...837..132V} {837, 132}

\bibitem[\protect\citeauthoryear{{van't Hoff} et~al.,}{{van't Hoff}
  et~al.}{2017}]{vant-hoff:2017}
{van't Hoff} M.~L.~R.,  et~al., 2017, \mn@doi [\aap]
  {10.1051/0004-6361/201629452}, \href
  {http://adsabs.harvard.edu/abs/2017A%26A...599A.101V} {599, A101}

\makeatother
\end{thebibliography}
\label{lastpage}

\end{document}